\documentclass[12pt]{article}
\usepackage{myart}

\renewcommand{\theequation}{\arabic{section}.\arabic{equation}}
\normalbaselines
\input tcilatex.tex

\begin{document}

\author{Yuri A. Rylov}
\title{Repeated dynamic quantization}
\date{Institute for Problems in Mechanics, Russian Academy of Sciences \\
101-1,Vernadskii Ave., Moscow, 119526, Russia \\
email: rylov@ipmnet.ru\\
Web site: {$http://rsfq1.physics.sunysb.edu/\symbol{126}rylov/yrylov.htm$}\\
or mirror Web site: {$http://gasdyn-ipm.ipmnet.ru/\symbol{126}%
rylov/yrylov.htm$}}
\maketitle

\begin{abstract}
In conventional quantum mechanics the quantum particle is a special object,
whose properties are described by special concepts and quantum principles.
The quantization is a special procedure, which is accompanied by
introduction of special concepts, and this procedure cannot be repeated. In
the model conception of quantum phenomena (MCQP) the quantum particle is a
partial case of a stochastic particle, and the quantum dynamics is a special
case of the stochastic particle dynamics. In MCQP the quantization is a
dynamical procedure, where a special quantum term is added to the Lagrangian
of the statistical ensemble of (classical) particles. This procedure can be
repeated many times, because it is not accompanied by introduction of new
concepts. It is very convenient from the formal viewpoint, because the set
of dynamic quantizations forms a one-parameter group, which allows one to
separate the dynamical and statistical (stochastic) properties.
\end{abstract}

\section{Introduction}

Sometimes investigation of a new class of physical phenomena is carried out
by two stages. At first, the simpler axiomatic conception based on simple
empiric considerations arises. Next, the axiomatic conception is replaced by
a more developed model conception, where axioms of the first stage are
obtained as properties of the model. Theory of thermal phenomena was
developed according to this scheme. At first, the thermodynamics (axiomatic
conception of thermal phenomena) appeared. Next, the statistical physics
(model conception of thermal phenomena) was developed. Axioms of
thermodynamics and properties of the caloric fluid were obtained as
properties of the chaotic molecular motion. A model conception offers few
advantages over the axiomatic conception. \textit{The investigation methods
and mathematical technique of the model conception are more subtle and
effective}, than those of the axiomatic conception. For instance, in the
framework of the statistical physics we can calculate the heat capacity and
other characteristics of the matter, whereas in the framework of
thermodynamics we can only measure them empirically. The second law of
thermodynamics is a postulate of thermodynamics. It is a formal evidence of
the axiomatic character of thermodynamics. In the framework of the
statistical physics the second law of thermodynamics is a corollary of the
heat model (chaotic molecular motion) and of the statistical principles.

The contemporary quantum theory is the first (axiomatic) stage in the
development of the microcosm physics. Formal evidences of this is an
existence of quantum principles, which are additional to primary principles
of classical physics. Appearance of the next (model) stage, where the
quantum principles are consequences of the model, seems to be unavoidable,
because only in this case the microcosm physics has a reliable foundation
for its further development. The model conception is attractive, because it
uses more subtle and effective mathematical methods of investigation.
Besides, it gives boundaries of the axiomatic conception application. We can
see this in example of interplay of statistical physics and thermodynamics.

Model conception of quantum phenomena (MCQP) looks as follows \cite{R2003a}.
As a result of a generalization of the Riemannian geometry we obtain a class
of such uniform and isotropic space-time geometries (nondegenerate
geometries \cite{R004,R01,R99}), where the free motion of particles is
primordially stochastic and the particle mass is geometrized. Experiments
show that the free motion of particles of small mass is stochastic, and this
stochasticity depends on the particle mass. Free motion of particles is to
be determined only by the space-time geometry, and nondegenerate geometries
with stochastic motion of free particle are more valid, than the Minkowski
space-time geometry with deterministic motion of free particles. It is
possible to choose such parameters of the nondegenerate space--time
geometry, that the statistical description of the stochastically moving
particles coincides with the quantum description (Schr\"{o}dinger equation).
The quantum constant $\hbar $ appears as an attribute of the space-time
geometry. The quantum principles appear to be corollaries of the
nonrelativistic approximation of the space-time model. Then the model
conception of quantum phenomena (MCQP) arises.

Briefly, MCQP can be formulated as follows. The space-time geometry depends
on the quantum constant $\hbar $ and generates the stochasticity, which
depends on the particle mass $m$ and on $\hbar $. Statistical ensemble of
stochastically moving particles is a kind of fluid, depending on $m$ and $%
\hbar $. Spin and wave function are attributes of any ideal fluid \cite{R99}%
. Describing the fluid (statistical ensemble) in terms of the wave function,
we obtain the quantum description (Schr\"{o}dinger equation).

Stochasticity of the free particle motion in the deterministic nondegenerate
space-time geometry is explained as follows. $\ $In the nondegenerate
geometry there are many vectors $\overrightarrow{P_{0}Q}$ of fixed length,
which are parallel to the vector $\overrightarrow{P_{0}P_{1}}$, whereas in
the degenerate geometry there is only one vector $\overrightarrow{P_{0}Q}$
of fixed length, which is parallel to the vector $\overrightarrow{P_{0}P_{1}}
$. The momentum vector $\overrightarrow{p}$ of a free particle is
transported in parallel along the particle world line. The momentum vector $%
\overrightarrow{p}$ is tangent to the world line and determines its
direction. In the degenerate geometry (for instance, in Minkowski geometry)
the momentum vector $\overrightarrow{p}$, transported in parallel, is
unique, and the world line is determined uniquely. In the nondegenerate
geometry there are many momentum vectors, transported in parallel, and the
world line is not determined uniquely. In other words, the world line
appears to be stochastic (random).

Replacement of the axiomatic conception of quantum phenomena (ACQP) by MCQP
is a rather radical modification of the existing theory of microcosm, and
for a correct evaluation of interplay between ACQP and MCQP one should take
into account such a factor as the style of investigation, which is usually
very conservative. Style of investigations influences strongly on the
evaluations of the achievements of a theory. There are two styles of
investigations: (1) classical style of investigations (C-style) and (2)
pragmatic style of investigations (P-style).

The C-style is a deductive style of investigations. It is very sensitive to
validity of the fundamental statements and principles of the physical
conception. Deductive character of the C-style does not allow to produce
investigations with false principles or fundamental statements, because all
corollaries are deduced from the primary principles. The essence of the
C-style is expressed by the Newton's slogan: ''Hypotheses non fingo''. If
some primary principles are false, the C-style of investigations does not
work.

On the contrary, the P-style of investigations is insensitive to the
validity of the primary principles. P-style uses deduction slightly, and it
can effectively work, when not all primary principles are valid. The P-style
uses short logic (short logical chains). It cannot use long logical chains.
It is unreliable, because one cannot be sure in the validity of primary
principles. Explanation of experimental data is the unique criterion, used
by the P-style. P-style is adequate for solution of small physical problems
and for description of restricted cycle of physical phenomena. The P-style
is inadequate for construction of fundamental physical conception, because
for such a construction the primary principles are to be true. To explain
new unknown physical phenomena, P-style uses new additional suppositions
(hypotheses), which connect different physical phenomena and explain them in
terms of the new hypotheses. P-style is effective and useful for
determination of connection between different physical phenomena. P-style
was used effectively at construction of the quantum mechanics.

Some contemporary investigators consider the P-style as a new investigation
style of contemporary physics, playing off it against the old-fashioned
style of classical physics (C-style). In reality the P-style is not new.
Ptolemeus and his successors used P-style in explanation of celestial
phenomena. Primary principles of celestial mechanics of that time contained
a mistake (the Sun rotates around the Earth). Nevertheless, Ptolemeus
succeeded to explain and to predict correctly the planet motion, because he
compensated the invalid primary statement by additional suppositions. Such a
conception, where the invalid primary statement is compensated by additional
suppositions will be referred to as a compensating conception. Such a
compensating conception is suitable for a correct description of special
physical phenomena, but it is impossible to construct a well defined
fundamental theory in framework of the compensating theory, which uses
P-style. For instance, in framework of the Ptolemaic doctrine it was
impossible to discover the Newton's gravitation law. This discovery took
place more than a century ago after the mistake in interplay with the Sun
and the Earth had been corrected by Copernicus.

A detailed investigation of interplay between P-style and C-style has been
made in \cite{R002a}. Here we restrict ourselves by the following
statements. From viewpoint of P-style the only criterion of the quality of
the scientific conception is an explanation of experimental data. Such
categories as logical structure of conception and the number of additional
hypotheses are not taken into account. Let two different conceptions $A$ and 
$B$ explain the same experimental data. If the conception $A$ was posed
earlier, from viewpoint of P-style it has the advantage of the conception $B$%
. In this case the logic of the P-style adherents is very simple: we have
conception $A$ and we do not need another conception $B$, which cannot
explain new experiments. The P-style adherents does not interested in the
relative quality of the conceptions $A$ and $B$.

From viewpoint of C-style the well defined conception must not have
additional hypotheses at all. If the conception has additional hypotheses,
it is a compensating conception, which contains mistakes (delusions) in its
primary principles. In the framework of C-style the criterion of explanation
of experimental data is also valid, but a direct test of experimental data
is difficult, because of the deductive character of the well defined
conception. If the well defined conception appears in that time, when there
is a competitive compensating conception, one needs a long time to deduce
mathematical technique from primary principles and explain existing
experimental data on the basis of this technique. In the case of competition
between the Ptolemaic and Copernicus doctrines one needed more than a
century. To demonstrate capacities of MCQP one needs to predict a new
physical phenomenon on the basis of the mathematical technique, generated by
MCQP. It is difficult to say what time do we need for such a prediction,
because construction of mathematical technique of MCQP is a difficult
problem.

Idea of the quantum mechanics foundation as a statistical description of
stochastically moving particles is a very old idea, but one has failed to
realize this idea because of mathematical problems. There were three serious
problems in realization of this idea:

\begin{enumerate}
\item  Construction of the adequate deterministic space-time geometry with
stochastically moving particles

\item  Construction of dynamical conception of the statistical description,
where the concept of the probability is not used

\item  Description of ideal fluid in terms of the wave function.
\end{enumerate}

Necessity of solving these problems has been existing since the beginning of
the quantum mechanics creation. To explain stochastic motion of particles,
the stochastic space-time geometries were invented \cite{M51,B70,B71}, but
these stochastic geometries were fortified geometries, i.e. geometries with
additional structures, given in the space-time, whereas one needs to
describe the particle stochasticity in the framework of deterministic
physical geometry, without introducing any additional structures (which are
additional hypotheses). It was well known that the Schr\"{o}dinger equation
can be presented in the form of hydrodynamic equations for irrotational flow
of some fluid \cite{M26}, but nobody could describe the rotational flow of
ideal fluid in terms of the wave function, because to make this, one needs
to integrate hydrodynamic equation for the ideal fluid. As a result the wave
function remains to be a primary object. After description of the ideal
fluid in terms of wave function \cite{R99} the fluid becomes to by a primary
object and the wave function may be considered to be an attribute of the
fluid. Then quantum phenomena can be interpreted in hydrodynamic terms. The
mentioned problems have been solved respectively in the papers \cite
{R01,R004}, \cite{R002,R2002,R98} and \cite{R99}.\label{01}

It is characteristic and essential that the first two problems have been
solved very simply only after discovery and correction of mistakes
(delusions) in classical approach to these problems. From viewpoint of
P-style one may suggest exotic additional suppositions, but a search of
mistakes in foundation of geometry and in principles of statistical
description is a hopeless undertaking, something like a scientific heresy.
Only scientific dissidents can look for mistakes in classical conceptions.
Nevertheless, the delusions (mistakes) have been found. These mistakes were
not logical or mathematical. They were associative. The ancient Egyptians
believed that all rivers flows towards the North, because they knew only one
river the Nile, which flowed exactly towards the North. The ancient
Egyptians associate direction of the river flow with the direction in the
space, and it was an associative delusion, because the origin of this
delusion is an incorrect association.

Impossibility of solution of the first problem (construction of a
deterministic space-time geometry with primordially stochastic motion of
free particles)s is connected with an associative delusion. We believe that
the straight line (analog of the straight line) is a one-dimensional set of
points in any physical geometry, because we know only such geometries, where
this statement is valid. As far as any physical geometry is a result of
deformation of the proper Euclidean geometry \cite{R01,R004}, this
one-dimensionality of the straight imposes unwarranted constraint on
possible deformation of the Euclidean geometry and eliminates true
space-time geometry from the list of possible space-time geometries.

Impossibility of solution of the second problem (construction of dynamical 
conception of the
statistical description) is connected with another associative delusion. In
the statistical physics the statistical description is produced in terms of
distributions which are a kind of the probability density. On this basis
many researchers believe that any statistical description is a probabilistic
description and try to construct the statistical description of stochastic
world lines as a probabilistic statistical description. Association of the
statistical description with the probability theory is an associative
delusion. Sometimes such a statistical description is possible, but not
always. Dynamical conception of statistical description is possible always,
although it is not so informative as the probabilistic statistical
description.

Our strategy of the microcosm investigation is as follows. We look for
mistakes in the classical approach to the microcosm description, find the
mistakes and correct them. Such a strategy is a safe strategy, because the
mistakes should be found and corrected independently of whether or not the
correction of the mistakes helps us to explain experimental data.

ACQP and MCQP are different conception describing quantum phenomena, as well
as thermodynamics and statistical physics are different conceptions,
describing the thermal phenomena. There is some correspondence between
procedures and methods in ACQP and in MCQP. This correspondence is described
by the following scheme

$
\begin{array}{cc}
\begin{array}{c}
\\ 
\text{ACQP}
\end{array}
& \text{MCQP} \\ 
\begin{array}{c}
\text{1. Additional hypotheses are used} \\ 
\text{(QM principles) }
\end{array}
& 
\begin{array}{c}
\text{1. \textit{No additional hypotheses are used}}
\end{array}
\\ 
\begin{array}{c}
\text{2. One kind of measurement, as } \\ 
\text{far as only one statistical average } \\ 
\text{object }\left\langle \mathcal{S}\right\rangle \text{ is considered. It
is } \\ 
\text{referred to as a quantum system}
\end{array}
& 
\begin{array}{c}
\text{2. Two kinds of measurement, because } \\ 
\text{two kinds of objects (individual }\mathcal{S}_{\mathrm{st}}\text{ } \\ 
\text{and statistical average }\left\langle \mathcal{S}\right\rangle \text{)
are } \\ 
\text{considered}
\end{array}
\\ 
\begin{array}{c}
\text{3. Quantization: procedure on } \\ 
\text{the conceptual level:} \\ 
\mathbf{p}\rightarrow -i\hbar \mathbf{\nabla }\;\;\;\text{etc. }
\end{array}
& 
\begin{array}{c}
\text{3. Dynamic quantization: relativistic } \\ 
\text{procedure on the dynamic level} \\ 
m^{2}\rightarrow m_{\mathrm{eff}}^{2}=m^{2}+\frac{\hbar ^{2}}{c^{2}}\left(
\kappa _{l}\kappa ^{l}+\partial _{l}\kappa ^{l}\right) 
\end{array}
\\ 
\begin{array}{c}
\text{4. Transition to classical description:} \\ 
\text{procedure on conceptual level} \\ 
\hbar \rightarrow 0\qquad \psi \rightarrow \left( \mathbf{x},\mathbf{p}%
\right) 
\end{array}
& 
\begin{array}{c}
\text{4. Dynamic disquantization: relativistic} \\ 
\text{ procedure on dynamic level} \\ 
\partial ^{k}\rightarrow \frac{j^{k}j^{l}}{j_{s}j^{s}}\partial _{l}
\end{array}
\\ 
\begin{array}{c}
\text{5. Combination of nonrelativistic } \\ 
\text{quantum technique with } \\ 
\text{principles of relativity}
\end{array}
& 
\begin{array}{c}
\text{5. Consequent relativistic description } \\ 
\text{at all stages}
\end{array}
\\ 
\begin{array}{c}
\text{6. Interpretation in terms of wave} \\ 
\text{function }\psi 
\end{array}
& 
\begin{array}{c}
\text{6. Interpretation in terms of statistical} \\ 
\text{average world lines (WL)} \\ 
\frac{dx^{i}}{d\tau }=j^{i}\left( x\right) ,\;\;\; \\ 
j^{k}=-\frac{i\hbar }{2}\left( \psi ^{\ast }\partial ^{k}\psi -\partial
^{k}\psi ^{\ast }\cdot \psi \right) 
\end{array}
\end{array}
$

\bigskip 

Now the main goal of MCQP is further development of its mathematical
technique, which distinguishes from the corresponding technique of ACQP. In
the present paper we consider only properties of the quantization procedure.
In conventional quantum mechanics (ACQP) the quantization procedure is a
transition from the classical description of a particle to the quantum
description of the particle. In ACQP the quantization is conceptual
procedure, where concepts of classical physics (coordinate $\mathbf{x}$ and
momentum $\mathbf{p}$) are replaced by operators (\textbf{\ }$\mathbf{x}$
and $-i\hbar \mathbf{\nabla }$) and the new concept -- wave function $\psi $
appears. The dynamical equations for variables $\mathbf{x,p}$ are replaced
by the Schr\"{o}dinger (or Klein-Gordon) equation for the wave function $%
\psi $. In ACQP the wave function $\psi $ is a fundamental object of the
theory, and properties of $\psi $ are described by quantum principles.

In the case of MCQP the quantization is a dynamical procedure. No new
concepts appear. Dynamic equations for the statistical ensemble $\mathcal{E}%
_{\mathrm{d}}\left[ \mathcal{S}_{\mathrm{d}}\right] $ of deterministic
(classical) particles $\mathcal{S}_{\mathrm{d}}$ are transformed to the
dynamic equations for the statistical ensemble $\mathcal{E}_{\mathrm{st}}%
\left[ \mathcal{S}_{\mathrm{st}}\right] $ of stochastic (quantum) particles $%
\mathcal{S}_{\mathrm{st}}$. These dynamic equations are equivalent to the
Schr\"{o}dinger (or Klein-Gordon) equation for the wave function $\psi $,
which describes the state of this statistical ensemble $\mathcal{E}_{\mathrm{%
st}}\left[ \mathcal{S}_{\mathrm{st}}\right] $. In MCQP the wave function is
simply a method of description of the statistical ensemble, which is a kind
of fluid. The wave function may be used for description of the statistical
ensemble $\mathcal{E}_{\mathrm{d}}\left[ \mathcal{S}_{\mathrm{d}}\right] $
of deterministic (classical) particles $\mathcal{S}_{\mathrm{d}}$, as well
as for description of the statistical ensemble $\mathcal{E}_{\mathrm{st}}%
\left[ \mathcal{S}_{\mathrm{st}}\right] $ of stochastic particles $\mathcal{S%
}_{\mathrm{st}}$. In both cases the statistical ensemble is a fluidlike
dynamic system.

In the framework of MCQP the quantization procedure (dynamic quantization)
may be produced many times, because dynamic quantization is not accompanied
by an introduction of new concepts. Formally, the dynamic quantization is an
addition of an accessory term to the Lagrangian. This accessory term
introduces a new field $\kappa $, describing the stochastic component of the
stochastic particle $\mathcal{S}_{\mathrm{st}}$ motion. We may repeat the
dynamic quantization many times, introducing any time a new field $\kappa $.
It appears that any new field $\kappa $ coincides with the existing field $%
\kappa $ to within a factor, and the repeated quantization leads only to a
change of the $\kappa $-field intensity. The $\kappa $-field is a very
important field, because it is responsible for pair production, and the pair
production mechanism can be described in terms of the $\kappa $-field \cite
{R003}. Nothing of that kind  cannot be obtained in the framework of ACQP,
because in ACQP the quantization is a conceptual procedure, which can be
carried out only once. Besides, in ACQP the $\kappa $-field is a constituent
of the wave function. The $\kappa $-field cannot be separated from the wave
function, because the wave function is a fundamental concept of ACQP.
Mechanism of the pair production cannot be determined and described in the
framework of ACQP. In a like way in thermodynamics we cannot obtain any
information on the molecular structure of the matter, whereas we can do this
in the framework of the statistical physics. In other words, MCQP carries
out more detailed description of quantum objects. Such a description is
impossible in the framework of ACQP.\ 

Mathematical technique of MCQP distinguishes from that of ACQP as well as
the mathematical technique of statistical physics distinguishes from that of
thermodynamics. Now a development of mathematical technique of MCQP is the
main problem of the MCQP\ construction. In the present paper we investigate
mathematical properties of the repeated dynamic quantization. In the
framework of MCQP the quantization procedure manifests group properties, and
this may appear to be interesting and useful for further investigations.

\section{Dynamic quantization}

The action for the statistical ensemble $\mathcal{E}\left[ \mathcal{S}_{%
\mathrm{d}}\right] $ of deterministic free relativistic particles $\mathcal{S%
}_{\mathrm{d}}$ has the form

\begin{equation}
\mathcal{E}\left[ \mathcal{S}_{\mathrm{d}}\right] :\qquad \mathcal{A}\left[ x%
\right] =\int \left\{ -mc\sqrt{g_{ik}\dot{x}^{i}\dot{x}^{k}}\right\} d\tau d%
\mathbf{\xi ,\qquad }\dot{x}^{k}\equiv \frac{dx^{k}}{d\tau }  \label{r1.1}
\end{equation}
where coordinates $x=\left\{ x^{i}\left( \xi \right) \right\} ,$ \ $%
i=0,1,2,3 $, $\;\xi =\left\{ \tau ,\mathbf{\xi }\right\} =\left\{ \tau ,\xi
_{1},\xi _{2},\xi _{3}\right\} $ describe the particle position in the
space-time. Lagrangian coordinates $\mathbf{\xi =}\left\{ \xi _{1},\xi
_{2},\xi _{3}\right\} $ label particles of the statistical ensemble $%
\mathcal{E}\left[ \mathcal{S}_{\mathrm{d}}\right] $. Here and in what
follows a summation is produced over repeated Latin indices $(0-3)$. To
produce dynamic quantization and to obtain the action for the statistical
ensemble $\mathcal{E}_{\mathrm{st}}\left[ \mathcal{S}_{\mathrm{st}}\right] $
of stochastic particles, it should make the change 
\begin{equation}
m^{2}\rightarrow m^{2}+\frac{\hbar ^{2}}{c^{2}}\left( \kappa _{l}\kappa
^{l}+\partial _{l}\kappa ^{l}\right) ,\qquad \partial _{l}\equiv \frac{%
\partial }{\partial x^{l}}  \label{r1.2}
\end{equation}
where $\kappa =\left\{ \kappa ^{l}\left( x\right) \right\} $, $l=0,1,2,3$
are new dynamic variables, describing the mean intensity of the stochastic
component of the particle motion (this component is absent for deterministic
particles). Dynamic equations for the $\kappa $-field $\kappa ^{l}$ are
determined from the variational principle by means of a variation with
respect to $\kappa ^{l}$. After the change (\ref{r1.2}) the action (\ref
{r1.1}) turns into the action for the statistical ensemble $\mathcal{E}_{%
\mathrm{st}}\left[ \mathcal{S}_{\mathrm{st}}\right] $ 
\begin{equation}
\mathcal{E}_{\mathrm{st}}\left[ \mathcal{S}_{\mathrm{st}}\right] :\qquad 
\mathcal{A}\left[ x,\kappa \right] =\int \left\{ -mcK\sqrt{g_{ik}\dot{x}^{i}%
\dot{x}^{k}}\right\} d\tau d\mathbf{\xi },\qquad K=\sqrt{1+\lambda
^{2}\left( \kappa _{l}\kappa ^{l}+\partial _{l}\kappa ^{l}\right) }
\label{r1.3}
\end{equation}
where $\lambda =\hbar /mc$ is the Compton wave length of the particle.

Meaning of the change (\ref{r1.2}) becomes to be clear, if we write the
action (\ref{r1.3}) in the nonrelativistic approximation, when $g_{ik}=$diag$%
\left\{ c^{2},-1,-1,-1\right\} $, \ $c^{-2}\left( \kappa ^{0}\right) ^{2}\ll 
\mathbf{\kappa }^{2}$, and $c^{2}\left( \dot{x}^{0}\right) ^{2}\gg \mathbf{%
\dot{x}}^{2}$. We have in the nonrelativistic approximation instead of (\ref
{r1.3}) 
\begin{equation}
\mathcal{E}_{\mathrm{st}}\left[ \mathcal{S}_{\mathrm{st}}\right] :\qquad 
\mathcal{A}\left[ \mathbf{x},\mathbf{u}\right] =\int \left\{ -mc^{2}+\frac{m%
}{2}\left( \frac{d\mathbf{x}}{dt}\right) ^{2}+\frac{m}{2}\mathbf{u}^{2}-%
\frac{\hbar }{2}\mathbf{\nabla u}\right\} dtd\mathbf{\xi },  \label{r1.4}
\end{equation}
where $\mathbf{x}=\mathbf{x}\left( t,\mathbf{\xi }\right) ,\;\;\mathbf{u}=%
\mathbf{u}\left( t,\mathbf{x}\right) =\frac{\hbar }{m}\mathbf{\kappa }$. The
variable $\mathbf{u}$ describes the mean value of the stochastic component
of velocity. Energy $m\mathbf{u}^{2}/2$ associated with this stochastic
component is added to the energy associated with the regular velocity of the
particle. The last term in (\ref{r1.4}) describes connection between the
stochastic component of the velocity and the regular one.

Formally the change (\ref{r1.2}) with arbitrary parameter $a=\hbar ^{2}$%
\begin{equation}
m^{2}\rightarrow m^{2}+\frac{a}{c^{2}}\left( \kappa _{l}\kappa ^{l}+\partial
_{l}\kappa ^{l}\right)  \label{r1.4a}
\end{equation}
may be applied to the statistical ensemble (\ref{r1.3}) of stochastic
particles (deterministic particles are considered as stochastic ones with
vanishing stochasticity). Such a transformation changes the stochasticity
intensity, and we obtain the stochastic particle dynamics with other kind of
stochasticity. Such an approach allows one to obtain the stochastic particle
dynamics with continuous dependence on the stochasticity intensity,
described by the parameter $a=\hbar ^{2}$. Such a dependence on the
parameter allows one to separate dynamical properties from the statistical
properties, conditioned by the particle motion stochasticity. Of course,
results of description in the framework of ACQP depend also on the parameter 
$a=\hbar ^{2}$, but in this case a change of the parameter $a=\hbar ^{2}$
generates a change of quantum principles, which contains the parameter $%
a=\hbar ^{2}$. Besides, setting $a=\hbar ^{2}=0$ in the conventional quantum
description, we do not obtain the classical description, because in ACQP the
quantum description do not turn to the classical one at $\hbar \rightarrow 0$%
. In the conventional quantum description the dynamics is mixed with the
stochasticity in such a way, that separation of them is not a simple
problem. Mathematical reason of this tangle will be shown below.

Dynamic equation for the variables $\kappa ^{l}$ are obtained from the
action (\ref{r1.3}) by means of variation with respect to $\kappa ^{l}$%
\begin{equation}
\frac{\delta \mathcal{A}}{\delta \kappa ^{l}}=-\lambda ^{2}\kappa _{l}\frac{%
mcR}{K}+\lambda ^{2}\partial _{l}\frac{mcR}{2K}=0,\qquad l=0,1,2,3
\label{r1.5}
\end{equation}
where 
\begin{equation}
R=J\sqrt{g_{ik}\dot{x}^{i}\dot{x}^{k}},\qquad J=\frac{\partial \left( \tau
,\xi _{1},\xi _{2},\xi _{3}\right) }{\partial \left(
x^{0},x^{1},x^{2},x^{3}\right) }  \label{r1.6}
\end{equation}

Solution of equations (\ref{r1.5}) has the form 
\begin{equation}
\kappa _{l}=\partial _{l}\kappa ,\qquad l=0,1,2,3,\qquad \kappa =\frac{1}{2}%
\ln \frac{mcR}{K}  \label{r1.7}
\end{equation}

After a series of changes of variables and some integration the action (\ref
{r1.3}) is reduced to the form (See mathematical details in Appendix) 
\begin{equation}
\mathcal{A}\left[ \psi ,\psi ^{\ast }\right] =\int \left\{ b_{0}^{2}\partial
_{k}\psi ^{\ast }\partial ^{k}\psi -m^{2}c^{2}\rho -\frac{b_{0}^{2}}{4}%
\left( \partial _{l}s_{\alpha }\right) \left( \partial ^{l}s_{\alpha
}\right) \rho +\left( \hbar ^{2}-b_{0}^{2}\right) \frac{\partial _{l}\rho
\partial ^{l}\rho }{4\rho }\right\} d^{4}x  \label{r1.7a}
\end{equation}
where $\psi =\psi \left( x\right) $ is the two-component complex wave
function, and $\psi ^{\ast }$ is the quantity complex conjugate to $\psi $%
\begin{equation}
\psi =\left( _{\psi _{2}}^{\psi _{1}}\right) ,\qquad \psi ^{\ast }=\left(
\psi _{1}^{\ast },\psi _{2}^{\ast }\right) ,  \label{r1.9}
\end{equation}
\begin{equation}
\rho =\psi ^{\ast }\psi ,\qquad s_{\alpha }=\frac{\psi ^{\ast }\sigma
_{\alpha }\psi }{\rho },\qquad \alpha =1,2,3  \label{r1.10}
\end{equation}
where $\mathbf{\sigma }=\left\{ \sigma _{1},\sigma _{2},\sigma _{3}\right\} $
are the Pauli matrices. The quantity $b_{0}$ is an arbitrary real constant ($%
b_{0}\neq 0$). Here and in what follows, a summation is produced over
repeated Greek indices $(1-3)$. The dynamic system, described by the action,
is an ideal fluid, where the 4-current $j^{i}$ is described by the relation 
\begin{equation}
j^{l}=\frac{ib_{0}}{2}\left( \psi ^{\ast }\partial ^{l}\psi -\partial
^{l}\psi ^{\ast }\cdot \psi \right)  \label{r1.12}
\end{equation}
The quantities $s_{\alpha }$ describe vorticity of the fluid flow. If $%
s_{\alpha }=$const, $\alpha =1,2,3$, the fluid flow is irrotational.

In the case of the irrotational flow the action we should use linear
dependent components of the wave function $\psi _{1}=a\psi _{2}$,\ $a=$%
const. In this case $s_{\alpha }=$const, $\alpha =1,2,3$ and the action (\ref
{r1.7a}) is reduced to the form 
\begin{equation}
\mathcal{A}\left[ \psi ,\psi ^{\ast }\right] =\int \left\{ b_{0}^{2}\partial
_{k}\psi ^{\ast }\partial ^{k}\psi -m^{2}c^{2}\rho +\left( \hbar
^{2}-b_{0}^{2}\right) \frac{\partial _{l}\rho \partial ^{l}\rho }{4\rho }%
\right\} d^{4}x  \label{r1.8}
\end{equation}
Setting $b_{0}=\hbar $ in (\ref{r1.8}), we obtain the action for the
Klein-Gordon equation 
\begin{equation}
\mathcal{E}_{\mathrm{st}}\left[ \mathcal{S}_{\mathrm{st}}\right] :\qquad 
\mathcal{A}\left[ \psi ,\psi ^{\ast }\right] =\int \left\{ \hbar
^{2}\partial _{k}\psi ^{\ast }\partial ^{k}\psi -m^{2}c^{2}\psi ^{\ast }\psi
\right\} d^{4}x  \label{r1.14}
\end{equation}

Thus, the change (\ref{r1.2}) realizes quantization of dynamic equations for
a free relativistic particle by means of dynamic methods, i.e. without a
reference to the quantum principles.

In the action (\ref{r1.8}) $b_{0}$ is an arbitrary constant, and the actions
(\ref{r1.8}) and (\ref{r1.14}) describe the same dynamic system for any
value of $b_{0}\neq 0$. But there is a difference in description of the
statistical ensemble in terms of actions (\ref{r1.8}) and (\ref{r1.14}). The
dynamic equation generated by the action (\ref{r1.14}) is linear, whereas
the dynamic equation generated by the action (\ref{r1.8}) is linear only at $%
b_{0}^{2}=\hbar ^{2}$. On the other hand, if we set $\hbar =0$ in the action
(\ref{r1.8}), we obtain the classical description, whereas if we set $\hbar
=0$ in the action (\ref{r1.14}), we obtain no description at all. The fact
is that the constant $b_{0}$ is connected with dynamics, whereas the
constant $\hbar $ is connected with stochasticity. If we set $b_{0}=0$ in
the action (\ref{r1.8}), we suppress the dynamics. If we set $\hbar =0$ in
the action (\ref{r1.8}), we suppress the stochasticity. In the action (\ref
{r1.14}) $b_{0}=\hbar $, and setting $\hbar =0$ in the action (\ref{r1.14}),
we suppress stochasticity and dynamics simultaneously. Thus, in (\ref{r1.14}%
) the dynamics is mixed with the stochasticity, and this mixture is a
necessary condition of the dynamic equation linearity. A linearity of
dynamic equation is very attractive. ACQP considers this linearity as a
principle. The tangle of stochasticity and dynamics is a payment for this
linearity.

Let us apply the repeated dynamic quantization to the action (\ref{r1.3}).
We obtain instead of (\ref{r1.3})

\begin{equation}
\mathcal{A}\left[ x,\kappa _{\left( 1\right) },\kappa _{\left( 2\right) }%
\right] =\int \left\{ -mcK\sqrt{g_{ik}\dot{x}^{i}\dot{x}^{k}}\right\} d\tau d%
\mathbf{\xi }  \label{r1.14a}
\end{equation}
where now 
\begin{equation}
K=\sqrt{1+\lambda ^{2}\sum\limits_{A=1,2}\left( \kappa _{\left( A\right)
l}\kappa _{\left( A\right) }^{l}+\partial _{l}\kappa _{\left( A\right)
}^{l}\right) },  \label{r1.14b}
\end{equation}
Dynamic equations for $\kappa _{\left( A\right) l}$ have the form 
\begin{equation}
\frac{\delta \mathcal{A}}{\delta \kappa _{\left( A\right) }^{l}}=-\lambda
^{2}\kappa _{\left( A\right) l}\frac{mcR}{K}+\lambda ^{2}\partial _{l}\frac{%
mcR}{2K}=0,\qquad A=1,2  \label{r1.14c}
\end{equation}
\[
R=\sqrt{g_{ik}\dot{x}^{i}\dot{x}^{k}}\frac{\partial \left( \tau ,\xi
_{1},\xi _{2},\xi _{3}\right) }{\partial \left(
x^{0},x^{1},x^{2},x^{3}\right) }
\]

Solution of dynamic equations (\ref{r1.14c}) gives 
\begin{equation}
\kappa _{\left( 1\right) l}=\kappa _{\left( 2\right) l}=\frac{1}{2}\partial
_{l}\kappa ,\qquad \kappa =\ln \frac{mcR}{K}  \label{r1.14d}
\end{equation}
Substitution of (\ref{r1.14d}) in (\ref{r1.14b}) leads to 
\begin{equation}
K=\sqrt{1+\lambda ^{\prime 2}\left( \kappa _{l}\kappa ^{l}+\partial
_{l}\kappa ^{l}\right) },\qquad \lambda ^{\prime 2}=2\lambda ^{2}=2\left( 
\frac{\hbar }{mc}\right) ^{2}  \label{r1.14e}
\end{equation}

Comparing (\ref{r1.14c}) with (\ref{r1.3}), we conclude that two subsequent
dynamic quantizations with intensity described by the parameter $\hbar ^{2}$
are equivalent to one dynamic quantization with the intensity described by
the parameter $\hbar ^{\prime 2}=2\hbar ^{2}$.

Dynamic quantization does not depend on the form of the action
representation. For instance, let us apply the repeated dynamic quantization
to the action (\ref{r1.8}). Using replacement (\ref{r1.2}) in the action (%
\ref{r1.8}), we obtain additional term $\mathcal{A}_{\mathrm{add}}\left[
\psi ,\psi ^{\ast },\kappa \right] $ in the action (\ref{r1.8}) 
\begin{equation}
\mathcal{A}_{\mathrm{add}}\left[ \psi ,\psi ^{\ast },\kappa \right] =-\int
\left\{ \hbar ^{2}\left( \kappa _{l}\kappa ^{l}+\partial _{l}\kappa
^{l}\right) \psi ^{\ast }\psi \right\} d^{4}x  \label{r1.15}
\end{equation}
Dynamic equation for the $\kappa $-field have the form 
\begin{equation}
\frac{\delta \mathcal{A}}{\delta \kappa ^{l}}+\frac{\delta \mathcal{A}_{%
\mathrm{add}}}{\delta \kappa ^{l}}=\frac{\delta \mathcal{A}_{\mathrm{add}}}{%
\delta \kappa ^{l}}=-2\hbar ^{2}\kappa _{l}\left( \psi ^{\ast }\psi \right)
+\hbar ^{2}\partial _{l}\left( \psi ^{\ast }\psi \right) =0,\qquad l=0,1,2,3
\label{r1.16}
\end{equation}
Solution of dynamic equations (\ref{r1.16}) can be written in the form 
\begin{equation}
\kappa _{l}\equiv \frac{1}{2}\partial _{l}\ln \rho =\frac{1}{2}\partial
_{l}\ln \left( \psi ^{\ast }\psi \right)  \label{r1.17}
\end{equation}

After substitution of (\ref{r1.17}) in (\ref{r1.15}) we obtain 
\begin{equation}
\mathcal{A}_{\mathrm{add}}\left[ \psi ,\psi ^{\ast }\right] =-\int \hbar
^{2}\left( \frac{\partial _{l}\rho \partial ^{l}\rho }{4\rho }-\frac{%
\partial _{l}\rho \partial ^{l}\rho }{2\rho }+\frac{1}{2}\partial
_{l}\partial ^{l}\rho \right) d^{4}x  \label{r1.18}
\end{equation}
The last term in (\ref{r1.18}) has the form of divergence. It does not
contribute to dynamic equations and can be omitted. Uniting (\ref{r1.8}) and
(\ref{r1.18}), we obtain 
\begin{eqnarray}
\mathcal{A}\left[ \psi ,\psi ^{\ast }\right] +\mathcal{A}_{\mathrm{add}}%
\left[ \psi ,\psi ^{\ast }\right] &=&\int \left\{ b_{0}^{2}\partial _{k}\psi
^{\ast }\partial ^{k}\psi -\frac{b_{0}^{2}}{4}\left( \partial _{l}s_{\alpha
}\right) \left( \partial ^{l}s_{\alpha }\right) \rho \right.  \nonumber \\
&&\left. -m^{2}c^{2}\rho +\left( 2\hbar ^{2}-b_{0}^{2}\right) \frac{\partial
_{l}\rho \partial ^{l}\rho }{4\rho }\right\} d^{4}x  \label{r1.19}
\end{eqnarray}
The action (\ref{r1.19}), obtained as a result of the repeated dynamic
quantization, distinguishes from the action (\ref{r1.8}) only in the sense
that the quantum constant $\hbar $ is replaced by the quantum constant $%
\hbar ^{\prime }=\sqrt{2}\hbar $.

\section{Discussion}

The repeated dynamic quantization manifests the difference between the
approach of ACQP and that of MCQP. This difference lies mainly in the
interpretation of the $\kappa $-field. From the viewpoint of ACQP the $%
\kappa $-field does not exist at all, because according to (\ref{r1.17}) it
is a constituent of the wave function, and the wave function is an attribute
of the particle. In the framework of ACQP there is no necessity to consider
the $\kappa $-field, it is sufficient to consider the corresponding wave
function. In ACQP the wave function is a fundamental object of ACQP, whose
properties are described by the quantum axiomatics, and it is useless to
divide the wave function into its constituents.

In MCQP the wave function is only a method of description of the statistical
ensemble $\mathcal{E}_{\mathrm{st}}\left[ \mathcal{S}_{\mathrm{st}}\right] $%
, consisting of stochastic particles $\mathcal{S}_{\mathrm{st}}$. Regular
component of the stochastic particle motion is described by the 4-current $%
j^{k}$, whereas the stochastic component is described by the $\kappa $-field 
$\kappa ^{l}$. From formal viewpoint the $\kappa $-field is a relativistic
force field, which is generated by the regular component of motion, and
which can exist separately from its source \cite{R003}. The $\kappa $-field
interacts with regular component of the particle motion. Two different
stochastic particle can interact via their common $\kappa $-field in a like
way, as two charged particles interact via their common electromagnetic
field. Interaction of two particles via the $\kappa $-field takes place only
in MCQP. This property is absent in ACQP, and it is a serious defect of
ACQP, because the $\kappa $-field can produce the particle-antiparticle
pairs. Neither electromagnetic field, nor gravitational one can produce
pairs, because they do not change the particle mass, that is necessary for
the pair production. Only $\kappa $-field can produce pairs, because the
factor $K$ in (\ref{r1.3}) can make the particle mass $m$ to be imaginary,
when $K^{2}<0$. It is necessary for the particle 4-velocity component $%
dx^{0}/d\tau $ can change its sign.

The pair production effect is the crucial effect of the high energy particle
collision. Experiments show that the pair production is an essentially
quantum effect. Now there is no satisfactory mechanism of the pair
production. Apparently, this mechanism is connected with application of the $%
\kappa $-field. At any rate, the pair production by means of the given
time-dependent $\kappa $-field is obtained \cite{R003}, whereas the pair
production at the collision of two relativistic particles is an unsolved
problem. The conventional description of the pair production in the
framework of the quantum field theory is unsatisfactory in some aspects (See
for details \cite{R003}).

\appendix
\renewcommand{\theequation}{\Alph{section}.\arabic{equation}}
\renewcommand{\thesection}{Appendix \Alph{section}.}
\section{Transformation of the action}

Let us transform the action (\ref{r1.3}) to the form (\ref{r1.8}). Instead
of $\tau $ we introduce  the variable $\xi _{0}$, and rewrite (\ref{r1.3})
in the form 
\begin{equation}
\qquad \mathcal{A}\left[ x,\kappa \right] =\int \left\{ -mcK\sqrt{g_{ik}\dot{%
x}^{i}\dot{x}^{k}}\right\} d^{4}\xi ,\qquad K=\sqrt{1+\lambda ^{2}\left(
\kappa _{l}\kappa ^{l}+\partial _{l}\kappa ^{l}\right) }  \label{A.1}
\end{equation}
where $\xi =\left\{ \xi _{0},\mathbf{\xi }\right\} =\left\{ \xi _{k}\right\} 
$,\ \ $k=0,1,2,3$, $x=\left\{ x^{k}\left( \xi \right) \right\} $,\ \ $%
k=0,1,2,3$.

Let us consider variables $\xi =\xi \left( x\right) $ in (\ref{A.1}) as
dependent variables and variables $x$ as independent variables. Let the
Jacobian 
\begin{equation}
J=\frac{\partial \left( \xi _{0},\xi _{1},\xi _{2},\xi _{3}\right) }{%
\partial \left( x^{0},x^{1},x^{2},x^{3}\right) }=\det \left| \left| \xi
_{i,k}\right| \right| ,\qquad \xi _{i,k}\equiv \partial _{k}\xi _{i},\qquad
i,k=0,1,2,3  \label{A.3}
\end{equation}
be considered to be a multilinear function of $\xi _{i,k}$. Then 
\begin{equation}
d^{4}\xi =Jd^{4}x,\qquad \dot{x}^{i}\equiv \frac{dx^{i}}{d\xi _{0}}\equiv 
\frac{\partial \left( x^{i},\xi _{1},\xi _{2},\xi _{3}\right) }{\partial
\left( \xi _{0},\xi _{1},\xi _{2},\xi _{3}\right) }=J^{-1}\frac{\partial J}{%
\partial \xi _{0,i}},\qquad i=0,1,2,3  \label{A.4}
\end{equation}
After transformation to dependent variables $\xi $ the action (\ref{A.1})
takes the form 
\begin{equation}
\mathcal{A}\left[ \xi ,\kappa \right] =-\int mcK\sqrt{g_{ik}\frac{\partial J%
}{\partial \xi _{0,i}}\frac{\partial J}{\partial \xi _{0,k}}}d^{4}x\mathbf{,}
\label{A.5}
\end{equation}

We introduce new variables 
\begin{equation}
j^{k}=\frac{\partial J}{\partial \xi _{0,k}},\qquad k=0,1,2,3  \label{A.6}
\end{equation}
by means of Lagrange multipliers $p_{k}$%
\begin{equation}
\mathcal{A}\left[ \xi ,\kappa ,j,p\right] =\int \left\{ -mcK\sqrt{%
g_{ik}j^{i}j^{k}}+p_{k}\left( \frac{\partial J}{\partial \xi _{0,k}}%
-j^{k}\right) \right\} d^{4}x\mathbf{,}  \label{A.7}
\end{equation}
Variation with respect to $\xi _{i}$ gives 
\begin{equation}
\frac{\delta \mathcal{A}}{\delta \xi _{i}}=-\partial _{l}\left( p_{k}\frac{%
\partial ^{2}J}{\partial \xi _{0,k}\partial \xi _{i,l}}\right) =0,\qquad
i=0,1,2,3  \label{A.8}
\end{equation}
Using identities 
\begin{equation}
\frac{\partial ^{2}J}{\partial \xi _{0,k}\partial \xi _{i,l}}\equiv
J^{-1}\left( \frac{\partial J}{\partial \xi _{0,k}}\frac{\partial J}{%
\partial \xi _{i,l}}-\frac{\partial J}{\partial \xi _{0,l}}\frac{\partial J}{%
\partial \xi _{i,k}}\right)   \label{A.9}
\end{equation}
\begin{equation}
\frac{\partial J}{\partial \xi _{i,l}}\xi _{k,l}\equiv J\delta
_{k}^{i},\qquad \partial _{l}\frac{\partial ^{2}J}{\partial \xi
_{0,k}\partial \xi _{i,l}}\equiv 0  \label{A.10}
\end{equation}
one can test by direct substitution that the general solution of linear
equations (\ref{A.8}) has the form 
\begin{equation}
p_{k}=b_{0}\left( \partial _{k}\varphi +g^{\alpha }\left( \mathbf{\xi }%
\right) \partial _{k}\xi _{\alpha }\right) ,\qquad k=0,1,2,3  \label{A.11}
\end{equation}
where $b_{0}\neq 0$ is a constant, $g^{\alpha }\left( \mathbf{\xi }\right)
,\;\;\alpha =1,2,3$ are arbitrary functions of $\mathbf{\xi =}\left\{ \xi
_{1},\xi _{2},\xi _{3}\right\} $, and $\varphi $ is the dynamic variable $%
\xi _{0}$, which ceases to be fictitious. Let us substitute (\ref{A.11}) in (%
\ref{A.7}). The term of the form $\partial _{k}\varphi \partial J/\partial
\xi _{0,k}$ is reduced to Jacobian and does not contribute to dynamic
equation. The terms of the form $\xi _{\alpha ,k}\partial J/\partial \xi
_{0,k}$ vanish due to identities (\ref{A.10}). We obtain 
\begin{equation}
\mathcal{A}\left[ \varphi ,\mathbf{\xi },\kappa ,j\right] =\int \left\{ -mcK%
\sqrt{g_{ik}j^{i}j^{k}}-j^{k}p_{k}\right\} d^{4}x\mathbf{,}  \label{A.12}
\end{equation}
where quantities $p_{k}$ are determined by the relations (\ref{A.11})

Variation of (\ref{A.12}) with respect to $\kappa ^{l}$ gives 
\begin{equation}
\frac{\delta \mathcal{A}}{\delta \kappa ^{l}}=-\frac{\lambda ^{2}mc\sqrt{%
g_{ik}j^{i}j^{k}}}{K}\kappa _{l}+\partial _{l}\frac{\lambda ^{2}mc\sqrt{%
g_{ik}j^{i}j^{k}}}{2K}=0  \label{A.12a}
\end{equation}
It can be written in the form 
\begin{equation}
\kappa _{l}=\partial _{l}\kappa =\frac{1}{2}\partial _{l}\ln \rho ,\qquad
e^{2\kappa }=\frac{\rho }{\rho _{0}}\equiv \frac{\sqrt{j_{s}j^{s}}}{\rho
_{0}mcK},  \label{A.13}
\end{equation}
where $\rho _{0}=$const is the integration constant. Substituting expression
for $K$ from (\ref{A.1}) in (\ref{A.13}), we obtain dynamic equation for $%
\kappa $%
\begin{equation}
\hbar ^{2}\left( \partial _{l}\kappa \cdot \partial ^{l}\kappa +\partial
_{l}\partial ^{l}\kappa \right) =\frac{e^{-4\kappa }j_{s}j^{s}}{\rho _{0}^{2}%
}-m^{2}c^{2}  \label{A.14}
\end{equation}

Variation of (\ref{A.12}) with respect to $j^{k}$ gives 
\begin{equation}
p_{k}=-\frac{mcKj_{k}}{\sqrt{g_{ls}j^{l}j^{s}}}  \label{A.16}
\end{equation}
or 
\begin{equation}
p_{k}g^{kl}p_{l}=m^{2}c^{2}K^{2}  \label{A.15}
\end{equation}
Substituting the second equation (\ref{A.13}) in (\ref{A.16}), we obtain 
\begin{equation}
j_{k}=-\rho _{0}e^{2\kappa }p_{k},  \label{A.17}
\end{equation}

Now we eliminate the variables $j^{k}$ from the action (\ref{A.12}), using
relation (\ref{A.17}) and (\ref{A.13}). We obtain 
\begin{equation}
\mathcal{A}\left[ \varphi ,\mathbf{\xi },\kappa \right] =\int \rho
_{0}e^{2\kappa }\left\{ -m^{2}c^{2}K^{2}+p^{k}p_{k}\right\} d^{4}x\mathbf{,}
\label{A.28}
\end{equation}
where $p_{k}$ is determined by the relation (\ref{A.11}). Using expression (%
\ref{A.1}) for $K,$ the first term of the action (\ref{A.28}) can be
transformed as follows. 
\begin{eqnarray*}
-m^{2}c^{2}e^{2\kappa }K^{2} &=&-m^{2}c^{2}e^{2\kappa }\left( 1+\lambda
^{2}\left( \partial _{l}\kappa \partial ^{l}\kappa +\partial _{l}\partial
^{l}\kappa \right) \right) \\
&=&-m^{2}c^{2}e^{2\kappa }+\hbar ^{2}e^{2\kappa }\partial _{l}\kappa
\partial ^{l}\kappa -\frac{\hbar ^{2}}{2}\partial _{l}\partial
^{l}e^{2\kappa }
\end{eqnarray*}

Let us take into account that the last term has the form of divergence. It
does not contribute to dynamic equations and can be omitted. Omitting this
term, we obtain 
\begin{equation}
\mathcal{A}\left[ \varphi ,\mathbf{\xi },\kappa \right] =\int \rho
_{0}e^{2\kappa }\left\{ -m^{2}c^{2}+\hbar ^{2}\partial _{l}\kappa \partial
^{l}\kappa +p^{k}p_{k}\right\} d^{4}x\mathbf{,}  \label{A.29}
\end{equation}

Instead of dynamic variables $\varphi ,\mathbf{\xi },\kappa $ we introduce $%
n $-component complex function 
\begin{equation}
\psi =\left\{ \psi _{\alpha }\right\} =\left\{ \sqrt{\rho }e^{i\varphi
}u_{\alpha }\left( \mathbf{\xi }\right) \right\} =\left\{ \sqrt{\rho _{0}}%
e^{\kappa +i\varphi }u_{\alpha }\left( \mathbf{\xi }\right) \right\} ,\qquad
\alpha =1,2,...n  \label{A.30}
\end{equation}
Here $u_{\alpha }$ are functions of only $\mathbf{\xi }=\left\{ \xi _{1},\xi
_{2},\xi _{3}\right\} $, having the following properties 
\begin{equation}
\sum\limits_{\alpha =1}^{\alpha =n}u_{\alpha }^{\ast }u_{\alpha }=1,\qquad -%
\frac{i}{2}\sum\limits_{\alpha =1}^{\alpha =n}\left( u_{\alpha }^{\ast }%
\frac{\partial u_{\alpha }}{\partial \xi _{\beta }}-\frac{\partial u_{\alpha
}^{\ast }}{\partial \xi _{\beta }}u_{\alpha }\right) =g^{\beta }\left( 
\mathbf{\xi }\right)  \label{A.31}
\end{equation}
where ($^{\ast }$) denotes complex conjugation. The number $n$ of components
of the wave function $\psi $ is chosen in such a way, that equations (\ref
{A.31}) have a solution. Then we obtain 
\begin{eqnarray}
\psi ^{\ast }\psi &\equiv &\sum\limits_{\alpha =1}^{\alpha =n}\psi _{\alpha
}^{\ast }\psi _{\alpha }=\rho =\rho _{0}e^{2\kappa },\qquad \partial
_{l}\kappa =\frac{\partial _{l}\left( \psi ^{\ast }\psi \right) }{2\psi
^{\ast }\psi }  \label{A.33} \\
p_{k} &=&-\frac{ib_{0}\left( \psi ^{\ast }\partial _{k}\psi -\partial
_{k}\psi ^{\ast }\cdot \psi \right) }{2\psi ^{\ast }\psi },\qquad k=0,1,2,3
\label{A.34}
\end{eqnarray}
Substituting relations (\ref{A.33}), (\ref{A.34}) in (\ref{A.29}), we obtain
the action, written in terms of the wave function $\psi $%
\begin{eqnarray}
\mathcal{A}\left[ \psi ,\psi ^{\ast }\right] &=&\int \left\{ \left[ \frac{%
ib_{0}\left( \psi ^{\ast }\partial _{k}\psi -\partial _{k}\psi ^{\ast }\cdot
\psi \right) }{2\psi ^{\ast }\psi }\right] \left[ \frac{ib_{0}\left( \psi
^{\ast }\partial ^{k}\psi -\partial ^{k}\psi ^{\ast }\cdot \psi \right) }{%
2\psi ^{\ast }\psi }\right] \right.  \nonumber \\
&&+\left. \hbar ^{2}\frac{\partial _{l}\left( \psi ^{\ast }\psi \right)
\partial ^{l}\left( \psi ^{\ast }\psi \right) }{4\left( \psi ^{\ast }\psi
\right) ^{2}}-m^{2}c^{2}\right\} \psi ^{\ast }\psi d^{4}x  \label{A.35}
\end{eqnarray}

Now we consider the case, when $n=2$, and the wave function has two
components. In this case 
\begin{equation}
\psi =\left( _{\psi _{2}}^{\psi _{1}}\right) ,\qquad \psi ^{\ast }=\left(
\psi _{1}^{\ast },\psi _{2}^{\ast }\right) ,  \label{A.43}
\end{equation}
and we have the following identity 
\begin{eqnarray}
&&\frac{\left( \psi ^{\ast }\partial _{l}\psi -\partial _{l}\psi ^{\ast
}\cdot \psi \right) \left( \psi ^{\ast }\partial ^{l}\psi -\partial ^{l}\psi
^{\ast }\cdot \psi \right) }{4\rho }-\frac{\left( \partial _{l}\rho \right)
\left( \partial ^{l}\rho \right) }{4\rho }  \nonumber \\
&\equiv &-\partial _{l}\psi ^{\ast }\partial ^{l}\psi +\frac{1}{4}\left(
\partial _{l}s_{\alpha }\right) \left( \partial ^{l}s_{\alpha }\right) \rho
\label{A.41}
\end{eqnarray}
where 3-vector $\mathbf{s=}\left\{ s_{1},s_{2},s_{3},\right\} $ is defined
by the relation 
\begin{equation}
\rho =\psi ^{\ast }\psi ,\qquad s_{\alpha }=\frac{\psi ^{\ast }\sigma
_{\alpha }\psi }{\rho },\qquad \alpha =1,2,3  \label{A.42}
\end{equation}
and Pauli matrices $\mathbf{\sigma }=\left\{ \sigma _{1},\sigma _{2},\sigma
_{3}\right\} $ have the form 
\begin{equation}
\sigma _{1}=\left( 
\begin{array}{cc}
0 & 1 \\ 
1 & 0
\end{array}
\right) ,\qquad \sigma _{2}=\left( 
\begin{array}{cc}
0 & -i \\ 
i & 0
\end{array}
\right) ,\qquad \sigma _{1}=\left( 
\begin{array}{cc}
1 & 0 \\ 
0 & -1
\end{array}
\right)  \label{A.44}
\end{equation}
Using identity (\ref{A.41}), we obtain from (\ref{A.35}) 
\begin{equation}
\mathcal{A}\left[ \psi ,\psi ^{\ast }\right] =\int \left\{ b_{0}^{2}\partial
_{k}\psi ^{\ast }\partial ^{k}\psi -m^{2}c^{2}\rho -\frac{b_{0}^{2}}{4}%
\left( \partial _{l}s_{\alpha }\right) \left( \partial ^{l}s_{\alpha
}\right) \rho +\left( \hbar ^{2}-b_{0}^{2}\right) \frac{\partial _{l}\rho
\partial ^{l}\rho }{4\rho }\right\} d^{4}x  \label{A.45}
\end{equation}
which coincide with (\ref{r1.7a}).


\begin{thebibliography}{99}
\bibitem{R2003a}  Yu. A. Rylov, Model conception of quantum phenomena:
logical structure and investigation methods. (Available at
http://arXiv.org/abs/physics/0310050, v2).

\bibitem{R004}  Yu. A. Rylov, Deformation principle as a foundation of
physical geometry. (Available at http://arXiv.org/abs/math.GM/0312160)

\bibitem{R01}  Yu.A. Rylov, Geometry without topology as a new conception of
geometry.\textit{\ Int. Jour. Mat. \& Mat. Sci.} \textbf{30}, iss. 12,
733-760, (2002), (available at http://arXiv.org/abs/math.MG/0103002).

\bibitem{R99}  Yu.A. Rylov, Spin and wave function as attributes of ideal
fluid. \textit{J. Math. Phys.} \textbf{40}, No.1, 256-278, (1999).

\bibitem{R002a}  Yu.A. Rylov Associative delusions and problem of their
overcoming. (Available at http://arXiv.org/abs/physics/0201065 ).

\bibitem{R2002}  Yu. A. Rylov, Dynamics of stochastic systems and
peculiarities of measurements in them. (In preparation, available at http://
arXiv.org /abs/physics /0210003).

\bibitem{M51}  K.~Menger, Probabilistic geometry. \textit{Proc. Nat. Acad.
Sci.,} \textbf{37}, 226-229, (1951).

\bibitem{B70}  D. I. Blokhintsev, \textit{Space and time in microcosm}.
Nauka, 1970. (in Russian).

\bibitem{B71}  D. I. Blokhintsev, Stochastic spaces. \textit{Fiz Elementar. 
\v{C}asti\v{c} i Atom Jadra} \textbf{5, }606-644, (1974). (in Russian).

\bibitem{M26}  E.~Madelung, Quanten theorie in hydrodynamischer Form. 
\textit{Z.Phys.} \textbf{40,} 322-326, (1926).

\bibitem{R002}  Yu.A. Rylov, Hamilton variational principle for statistical
ensemble of deterministic systems and its application for ensemble of
stochastic systems. \textit{Russ. J. Math. Phys.}, \textbf{9}, iss. 3,
335-348, (2002).

\bibitem{R98}  Yu.A. Rylov, Quantum mechanics as a dynamic construction. 
\textit{Found. Phys.} \textbf{28}, No.2, 245-271, (1998).

\bibitem{R003}  Yu. A. Rylov, Classical description of pair production.
(Available at http:// arXiv.org/abs/physics/0301020).
\end{thebibliography}
\end{document}